\documentclass[12pt]{article}
\usepackage{epsfig}

\newcommand{\bx}{{\bf x}}
\newcommand{\bv}{{\bf v}}
\newcommand{\BE}{\begin{equation}}
\newcommand{\EE}{\end{equation}}
\newcommand{\BA}{\begin{eqnarray}}
\newcommand{\EA}{\end{eqnarray}}

\begin{document}

\title{Sustained plankton blooms under open chaotic flows}
\author{Emilio Hern{\'a}ndez-Garc{\'\i}a and Crist\'obal L\'opez\\
Instituto Mediterraneo de Estudios Avanzados (IMEDEA)\\
CSIC-Universitat de les Illes Balears\\
E-07071 Palma de Mallorca, Spain}

\date{\small November 21, 2003}
\maketitle

\abstract{
We consider a predator-prey model of planktonic population dynamics,
of excitable character, living in an open and chaotic fluid flow, i.e.,
a state of fluid motion in which fluid trajectories are unbounded but a
chaotic region exists that is restricted to a localized area. Despite that excitability is a transient
phenomenon and that fluid trajectories are continuously
leaving the system, there is a regime of parameters where
the excitation remains permanently in the system, given rise
to a persistent plankton bloom. This regime is reached when the time scales associated 
to fluid stirring become slower than the ones associated to biological growth. 

\section{Introduction}

The impact of ocean hydrodynamic conditions on the biological activity of plankton species
has been a subject attracting the interest of researchers during several
decades \cite{hydroplank}. Most of the studies have addressed  the r{\^o}le
of vertical transport processes (advection, turbulent mixing, ...) 
to bring nutrients to the upper layers of the ocean, where most
of the plankton lives \cite{vertical}. It is now recognized that
horizontal (or lateral) transport mediated by mesoscale structures
such as eddies, jets, and fronts is also a major player in the dynamics
of plankton populations, providing the basic mechanism for patchiness
in the plankton distribution and influencing also key features
such as biological productivity \cite{patchs,productivity}. 

Ideas and methods borrowed from dynamical systems theory have been shown
to be a powerful tool to understand the basics of the influence of
mixing and transport on plankton dynamics, and on the more general
framework  of chemically or biologically interacting
species \cite{ActiveChaos}. They are complementary to the more
traditional {\it turbulence approach} \cite{hydroplank,OkuboPRSL}.
In particular, the concepts of chaotic advection \cite{Aref}
have been used to study fluid flow effects on models of populations
competing for resources \cite{competing}, or predator-prey 
dynamics of different kinds \cite{predatorprey}. 

Among the variety of flows considered by these approaches, 
{\sl open chaotic flows} appear singled out by the strong modifications
they introduce on chemical or biological processes occurring on
them \cite{openflows}. Briefly stated, an open chaotic flow is
a kind of fluid motion in which typical fluid particles only remain
for a finite time in the region of interest, but in such a way that
initially close fluid trajectories experience exponential separation
during that time (i.e. they undergo {\sl transient chaos} or
{\sl chaotic scattering} \cite{ChaoticScatt}). This situation
is very common in the wake of obstacles interposed on flows through
pipes or channels, or of islands on ocean currents. In the presence
of sufficiently strong recirculation regions, some nontypical fluid
trajectories become trapped forever in the region of interest,
forming a fractal set of zero measure called the {\sl chaotic saddle}.
On these trajectories, motion is permanently chaotic and can be characterized
by standard tools in nonlinear dynamics such as Lyapunov exponents, 
entropies, stable and unstable manifolds, etc. \cite{ChaoticScatt}.
Although they do not remain trapped forever, fluid trajectories close enough to the 
stable manifold of the chaotic saddle leave the system only after a long time. They do so 
along the unstable manifold of the chaotic saddle. Thus, if some 
marked particles (for example by a dye or by a particular temperature)
are transported by this type of flow, most of them will leave the
system soon and at long times only the ones close to the chaotic
saddle or its unstable manifold will be still present.  
This makes these filamental fractal structures observable
both in laboratory experiments
\cite{ExperimentSaddle} and in the Ocean \cite{Aristegui}. 

In this Paper we report numerical results on the behavior of a 
predator-prey population dynamics model of the {\sl excitable} 
type, modelling zooplankton-phytoplankton interaction, in which
the populations are transported by an open chaotic fluid flow. 
Excitability, in its simplest form, is the name given to the
behavior of dynamical systems with the following 
characteristics \cite{excitable,Murray}: a) they have a single fixed point, 
which is globally stable, as the only attractor of the dynamics;
 b) when the system is 
in the fixed point, small perturbations of this state return linearly
to the equilibrium, but perturbations above a threshold lead to a
large excursion (excitation-deexcitation cycle) in phase space, 
after which the system finally returns to the globally stable 
equilibrium point; c) there is a large separation between the 
time scales of excitation and deexcitation. Truscott and 
Brindley \cite{Brindleya,Brindleyb} showed that a large class 
of models of plankton dynamics involving two and three species 
display excitable dynamics in a realistic range of parameters. 
The essential ingredient defining the class was that the function 
describing predation of zooplankton on
phytoplankton should be of Hollings type III type. 

Realizations of excitable dynamics in stirred fluids have been 
considered in several contexts, including shear flows \cite{shear},
and closed chaotic flows \cite{PRLexcitable,GRL}. In \cite{PREexcitable}, 
the case of open flows was explicitly considered. The main finding was
the identification of a regime in which, despite the essential
transient character of the excitation cycle in excitable 
dynamics, and the transient nature of the chaotic scattering 
experienced by almost all trajectories in their passage 
through the observation region, a permanent pattern of 
excitation was established and remained within the system
for all times. The purpose of the present Paper is to 
check whether the phenomenon appears also in an excitable 
plankton dynamical model with realistic parameter values 
(an excitable FitzHugh-Nagumo dynamics was considered
in \cite{PREexcitable}, of no explicit relevance to plankton
modelling), thus supporting that excitability was the relevant 
ingredient, instead of the peculiarities of the particular 
chemical or biological interactions used. In addition, an 
oceanographically motivated flow will be used here, instead
of the rather academic {\sl blinking vortex-sink} model 
considered in \cite{PREexcitable}. As the main conclusion, 
we will see that the most relevant phenomenon, the development
of a permanently excited structure within the observation 
region, is found in the plankton model. Some details of its 
behavior, however, are dependent on geometrical properties of 
the particular flow used. 

\section{The plankton and the flow models}

A standard way to model chemical or biological activity
in fluid flows is in terms of advection-reaction-diffusion 
equations. The space- and time-dependent phytoplankton, $P=P(\bx,t)$, 
and zooplankton, $Z=Z(\bx,t)$, concentrations evolve according to
\BA
\frac{\partial}{\partial t}P+\bv\cdot\nabla P-D\nabla^2 P  &=&
r\left[ \beta P\left(1-\frac{P}{K}\right) - f\left(P\right) Z \right]
\nonumber \\
\frac{\partial}{\partial t}Z+\bv\cdot\nabla Z-D\nabla^2 Z  &=&
r\epsilon \left[  f\left(P\right) Z -\omega Z \right] \ .
\label{ard}
\EA
The left-hand-side terms represent the transport processes: both species 
are advected by the same fluid flow characterized by the velocity field
$\bv=\bv(\bx,t)$, that we assume to be incompressible. To focuss in the
r{\^o}le of horizontal transport, we choose $\bv$ to be two-dimensional, 
defined on a two-dimensional square  box $\Omega=[0,2L]\times[-L,L]$ that
constitutes our system domain, with Cartesian coordinates $\bx=(x,y)$.
As usual in ocean dynamics applications, small-scale complex turbulent 
motions not explicitly included in $\bv$ are modelled by an effective 
diffusion operator $D\nabla^2$ which is also the same for both species.
The right-hand-side, taken from \cite{Brindleya}, contains the 
biological interactions terms: $r$ controls the ratio of the transport
time scales to the biological activity time scales, $\epsilon$ sets
the ratio of phytoplankton- to the much larger zooplankton-growth
time scale, $K$ is the phytoplankton carrying capacity, $\beta$ 
the phytoplankton growth rate, $\omega$ a linear zooplankton mortality, 
and 
\BE
f(P)=\frac{P^2}{P_0^2+P^2}
\EE
is the Hollings type III response function, describing 
the zooplankton predation on phytoplankton. 

Following \cite{Brindleya} we take non-dimensional units such that $\beta=0.43$,$K=1$,
$\epsilon=0.01$, $P_0=0.053$, and $\sigma=0.34$. This corresponds to phytoplankton doubling times of the order of days, and zooplankton time scales in the range of months. Biological concentrations have been scaled so that the phytoplankton carrying capacity (of the order of 100 $\mu g$ of Nitrogen equivalent per liter) is the unity. For these and similar parameter values the biological dynamical system is in the excitable regime, and we study the influence of transport by varying its relative strength 
via the parameter $r$. $D$ is fixed to $10^{-5}$ and $L=9$, which means that the diffusive spatial scale corresponding to the phytoplankton doubling time is between three and four orders of magnitude smaller than system size. 

Since the velocity field $\bv=(v_x,v_y)$ is two-dimensional 
and incompressible it can be written in terms of a streamfunction
 $\Psi(x,y,t)$ that acts  formally as a Hamiltonian for the fluid trajectories: 
\BA
v_x &=& \frac{\partial\Psi}{\partial y}  \nonumber \\
v_y &=& -\frac{\partial\Psi}{\partial x} \ .
\label{hamilton}
\EA
We consider the following streamfunction:
\BE
\Psi=\Psi_0 \tanh\left( \frac{y}{w} \right)+\mu
 \exp\left(-\frac{(x-L)^2+y^2}{2\sigma^2}\right) \cos\left( k (y-vt) \right)\ .
\label{streamfunction}
\EE
It represents an oceanic jet perturbed by a localized 
wave-like feature, trapped by topography or some geographical
accident. The first term is the main jet, of width $w$, flowing
towards the East (i.e. towards the positive $x$ direction) with 
maximum velocity $\Psi_0/w$ at its center. The wave-like perturbation, 
of strength $\mu$, is represented by the second term. It is localized
in a region of size $\sigma$ around the point $(x,y)=(L,0)$, and the wavenumber 
and phase velocity (towards the North, or positive $y$ direction) 
are $k$ and $v$, respectively. The complete velocity field 
is time-periodic with period $2\pi/kv$.

Equations (\ref{hamilton}) and (\ref{streamfunction}) 
define a 1-degree of freedom time-periodic Hamiltonian dynamical
system for the fluid trajectories. Typically, this kind of system 
develops chaotic regions increasing in size as the strength of the perturbation, 
$\mu$, increases. But the region $\Omega$ is open with respect to
this flow, so that we have the situation of chaotic scattering: 
particles enter $\Omega$ from the West, following essentially straight
trajectories, experience transient chaos when reaching the 
wave region, and finally they leave the system.  
For $\mu$ large enough, recirculation gives birth to a chaotic
saddle in $\Omega$. We now study how the flow structures 
affect the plankton dynamics given by the right-hand-side of Eq. (\ref{ard}).

\section{Numerical results}

Equations (\ref{ard}) are solved by the semilagrangian method described
in \cite{PREexcitable}. The fixed point representing stable 
phytoplankton-zooplankton coexistence in the absence of flow 
and diffusion is given by $P=P_e$ and $Z=Z_e$, with $P_e=P_0\sqrt{\sigma/(1-\sigma)}=0.03827$ 
and $Z_e=\beta(1-P_e)(P_0^2+P_e^2)/P_e=0.04603$. We choose these values to be imposed
as Dirichlet boundary conditions on the boundary of $\Omega$. In this way fluid
particles enter in the system with a plankton content corresponding to the 
equilibrium concentrations, which is a rather natural condition from the
biological point of view. During an excitation phase, the values of phytoplankton concentration rise to $P\approx 0.8-0.9$

Since $(P_e,Z_e)$ is a stable equilibrium point, dynamics will be trivial 
without an initial seed to trigger the excitation dynamics. Our initial 
condition is a localized patch of high phytoplankton concentration close 
to the western part of $\Omega$:
$P(\bx,t=0)=P_e+Q\exp[-((x-x_0)^2+y^2)/l^2]$,
$Z(\bx,t=0)=Z_e$. 
We take $Q=0.5$, $x_0=0.3L$, and $l=0.11 L$. The jet transports the patch towards the scattering region,
where interesting dynamics occurs. The flow parameters are $w=1$, $\Psi_0=2$, $\sigma=2$, $k=1$, and $v=1$, giving a flow period $T=2\pi/kv=2\pi$. 

We first discuss the case with $\mu=0$, that is, no wave 
perturbation is present and the flow is simply a narrow 
jet without recirculations nor any chaotic behavior.
One of the curves (dashed-line) in  Fig. 
\ref{fig:phytot} shows the total density of phytoplankton (spatial average of $P$) within 
the domain $\Omega$ for this case, i.e. $\mu=0$,  and $r=10$.
 The initial patch initially grows, since 
diffusion converts it in the well known excitable propagating 
target wave, while the jet transports and deforms it. The patch however
reaches the Eastern boundary of $\Omega$ and leaves the system 
without any unexpected effect. This is what it is observed in the Figure:
the average concentration grows up in the short initial stage due to the advance of the excitation 
wave, and then decreases because the flow takes the excited
 material out of the system.

\begin{figure}
\epsfig{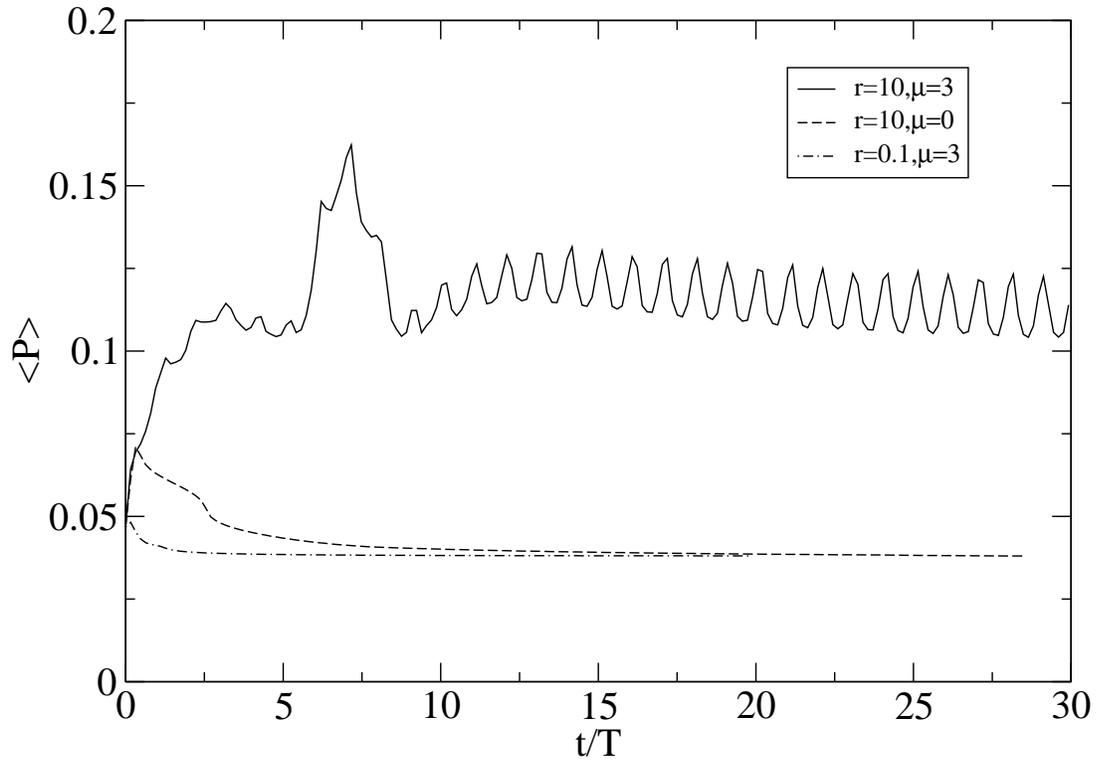}
\caption{Phytoplankton concentration, spatially averaged on $\Omega$, as a function of time (in units of the flow period $T=2\pi$). We plot three different cases: dashed-line is for $\mu=0$
and $r=10$, so that no chaotic scattering occurs; dashed-dotted line is for $\mu=3$  but $r$ small ($r=0.1$) corresponding to fast stirring, and the solid-line corresponds to a case of stirring slower than biological rates, i.e. $r$ large ($r=10$) and $\mu=3$, giving rise to a sustained excitation.
}
\label{fig:phytot}
\end{figure}

\begin{figure}
\epsfig{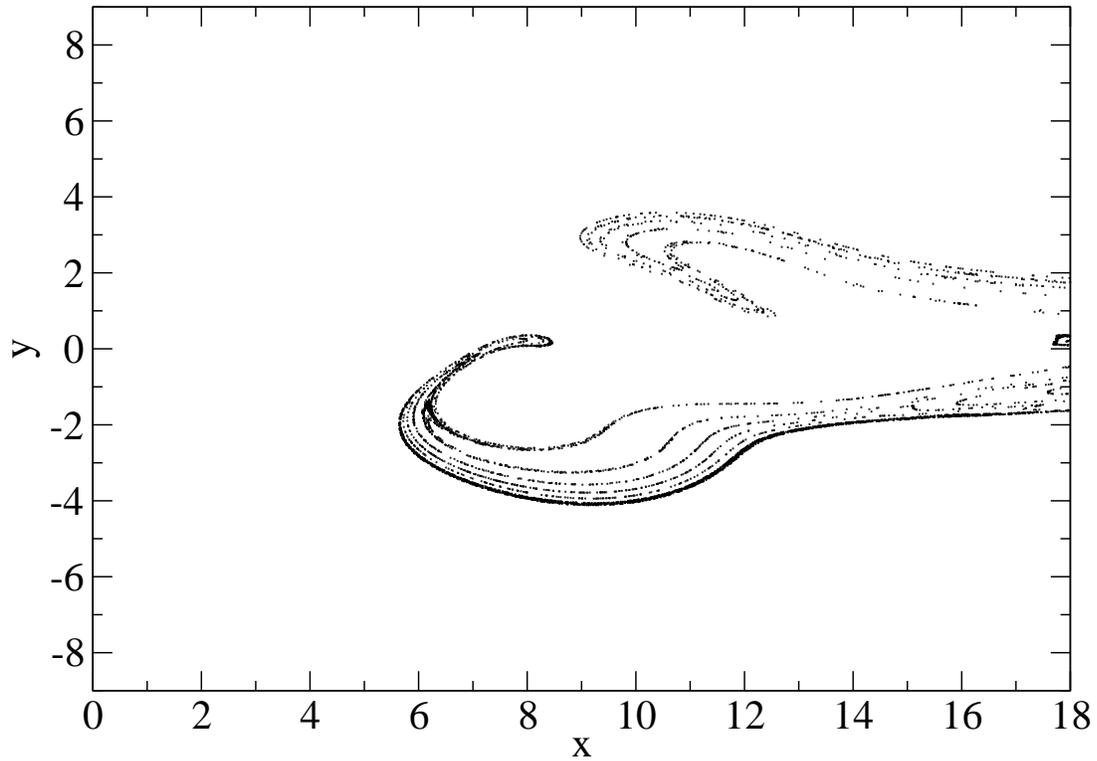}
\caption{The chaotic saddle and its unstable manifold for the flow defined by Eqs. (\ref{hamilton})-(\ref{streamfunction}). It is obtained by letting evolve during 6 periods of the flow 
a population of $30000$ particles, initially distributed in the Western part of the domain. After this time, $5366$ particles remain still in the system, tracing the saddle and its unstable manifold. 
}
\label{fig:saddle}
\end{figure}

We now consider $\mu=3$. A chaotic saddle is present in the 
system. Figure \ref{fig:saddle} shows the unstable manifold of this fractal set.
It has been visualized by releasing a large number of particles
with different initial conditions and letting them evolve with
the velocities given by (\ref{hamilton})-(\ref{streamfunction}). At sufficiently long 
times only the particles close to the saddle and its unstable
manifold remain in $\Omega$, which are the ones drawing the 
structure in Fig. \ref{fig:saddle}. Since the flow is 
time-periodic, saddle and manifolds change in time with 
the same period. Figure \ref{fig:saddle} gives the
configuration at a particular time (and the times obtained 
by adding one period, two periods, etc.). For small $r$, the biological dynamics is slow compared to the time scales of stirring by the flow. The 
phytoplankton patch is strongly deformed when reaching 
the scattering region. Plankton is stretched into long 
and thin filaments that become rapidly diluted into the surrounding
unexcited fluid by the effect of diffusion. Thus excitation is destroyed by the fast stirring. Increasing $r$, i.e. by making 
the biological dynamics faster or the flow slower, 
a dramatic change occurs. The transition to the new
regime occurs around $r \approx 1$. Plankton is again stretched in 
filaments but the width stabilizes and the excitation becomes
distributed in the system, without leaving it and oscillating in shape following the period of the flow. 
Some features of the distributions of both phytoplankton and zooplankton (see Fig. (\ref{fig:permanent})) seem to mimic the shape of the unstable manifold of the chaotic saddle, so that in these zones we can say that plankton is basically covering it with a finite width, but the correspondence is far from perfect, and some structures, as for example the excited filaments surrounding the whole scattering area, do not seem to be related to the saddle structure. This is the main result result of this Paper: 
{\sl transient} chaos plus {\sl transient} excitation give rise to a 
{\sl permanent} pattern of high biological activity (excitation) in the system.

\begin{figure}
\epsfig{figure=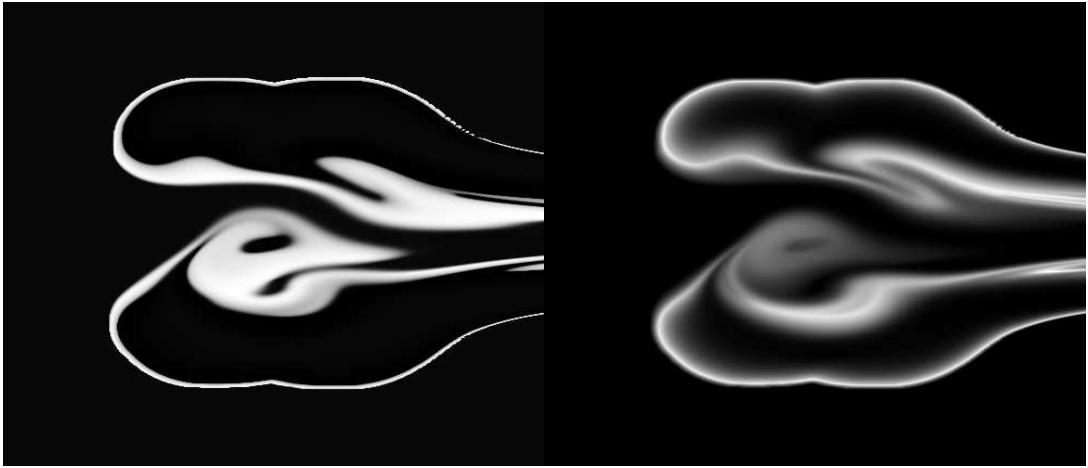,width=0.9\columnwidth,angle=0}
\caption{
Distribution of phytoplankton (left) and zooplankton (right) at time $t=100$ and parameters corresponding to the solid line in Fig. (\ref{fig:phytot}). Dark grey corresponds to low concentration and lighter grey to higher concentration. A state of permanent excitation is sustained in the region close to the chaotic saddle and its unstable manifold (being the rest of the domain in the unexcited equilibrium state). The shape of the distributions changes in time with the period of the flow. The images shown here are for the same phase of oscillation of the flow as in Fig. (\ref{fig:saddle}). 
}
\label{fig:permanent}
\end{figure}

\section{Discussion}

The reported result is of relevance in understanding the interplay between biology and fluid flow in aquatic biology. In particular it points a mechanism for enhanced productivity in the wake of islands, obstacles, or other perturbations on ocean currents. It adds to proposed mechanisms in closed flows \cite{productivity}. 

The explanation for the observed behavior can be elaborated along the lines of previous works \cite{PRLexcitable,PREexcitable,PhysicaA} as follows: The tendency of chaotic flow to stretch fluid elements into long and thin filaments competes with the effect of diffusion and biological growth, which tends to expand excited regions, so that a compensation can be achieved in some parameter range. Above a given biological growth rate, steady filament solutions appear, via a saddle-node bifurcation, in simplified models capturing some of the features of (\ref{ard}). This was demonstrated in \cite{PRLexcitable,PREexcitable,PhysicaA} for abstract excitable dynamics and in \cite{GRL} for model (\ref{ard}). Filament solutions appear also in other models of biological dynamics in fluid flows \cite{MartinEcolMod}. When this steady solution does not exist, initial perturbations decay as in the usual excitation-deexcitacion cycle, so that excitation disappears at long times. At sufficiently large biological growth rate, however, the steady filament solution exists, is stable, and the initial perturbation can be locally attracted by it. Chaotic flow deforms the filament solution obtained under simplified assumptions, but the results in  \cite{PRLexcitable,PREexcitable,GRL} indicate that it still provides a useful description of the process. Chaotic stretching and folding of the excited filament in a closed system ends up when it fills the whole domain, after which an homogeneous deexcitation finishes the excitation cycle. In an open system, however, the continuous outflow of excited material inhibits the filling of the full domain, so that distributions related to the simplified filament steady solutions can persist permanently. 

Despite that the previous results \cite{PRLexcitable,GRL,PREexcitable,PhysicaA} provide the above qualitative explanations for the phenomena reported in this Paper, there are a number of differences with respect to the previously considered open flow \cite{PREexcitable}. First, the spatial distribution does not follow the unstable manifold of the chaotic saddle as closely. Second, an additional transition to the disappearance of the permanent excitation was present in the blinking vortex-sink flow of \cite{PREexcitable} at sufficiently slow stirring or fast excitability. We have not found this second transition here despite having explored a rather broad parameter range. Our interpretation is that these differences arise from the different topology of the flows used in \cite{PREexcitable} and in this Paper. Here chaotic scattering tends to disperse fluid trajectories {\sl out} of the scattering region whereas the blinking vortex-sink flow of \cite{PREexcitable} attracts trajectories {\sl towards} the sinks inside the chaotic region. Thus it is natural to expect that the geometric structures inside the chaotic region will be more determinant in this last flow. Filament collisions, that we suspect to be responsible for the second transition in \cite{PREexcitable} are also more frequent in this last case. It would be convenient to provide additional quantitative support for these qualitative arguments. This will be the subject of future work.

\section*{Acknowledgements}

Financial support from MCyT (Spain) and FEDER through 
projects REN2001-0802-C02-01/MAR (IMAGEN) and BFM2000-1108 (CONOCE) is greatly acknowledged.


\end{document}